\documentclass{openjournal}
\usepackage{lipsum}
\usepackage{amsmath}

\usepackage{xcolor}
\usepackage{textgreek}
\usepackage[utf8]{inputenc}
\usepackage[english]{babel}

\usepackage{hyperref}
\hypersetup{
    unicode, 
    colorlinks=true,
    linkcolor=linkcolor,
    citecolor=linkcolor,
    filecolor=linkcolor,
    urlcolor=linkcolor,
}
\usepackage{color,colortbl}
\definecolor{linkcolor}{rgb}{0.0,0.3,0.5}
\usepackage{tensind}
\tensordelimiter{?}
\DeclareGraphicsExtensions{.bmp,.png,.jpg,.pdf}
\usepackage{verbatim}
\usepackage[normalem]{ulem}
\usepackage{orcidlink}
\usepackage{soul}

\graphicspath{ {./figs/} }

\newcommand{\kmsMpc}{\mathrm{km\,s^{-1}\,Mpc^{-1}}}

\begin{document}
\title{Selection effects in correlated observations with application to distance ladder observations}
\author{Will J. Percival\orcidlink{0000-0002-0644-5727}}
\email{will.percival@uwaterloo.ca}
\affiliation{Waterloo Centre for Astrophysics, University of Waterloo, Waterloo, ON N2L 3G1, Canada}
\affiliation{Department of Physics and Astronomy, University of Waterloo, Waterloo, ON N2L 3G1, Canada}
\affiliation{Perimeter Institute for Theoretical Physics, 31 Caroline St. North, Waterloo, ON NL2 2Y5, Canada}

\begin{abstract}
    For over a century, following the work of Eddington, Kapteyn, Malmquist and others, astronomers have wrestled with selection biases when making inferences from samples of objects. Typically, selection is performed on the same observations used to make the measurements of interest. However, selecting objects using one observable while analyzing another can also lead to a selection bias when the observables are correlated. Within a Bayesian framework, unmodelled selection effects correspond to a misspecified generative model. We derive the likelihood for truncated selection in correlated observables and demonstrate its usefulness by searching for residual selection effects of this form in recent distance ladder measurements of the Hubble constant $H_0$. We specifically look for the form of bias in Cepheid selection where the correction depends on the photometric uncertainties, which the data can constrain while simultaneously measuring $H_0$. We find mild evidence for non-zero residual selection corrections for Cepheids within calibrator galaxies at a significance of $1.3\sigma$ to $2.3\sigma$ depending on the distance-prior adopted, but find that allowing for them does potentially have an appreciable effect on $H_0$. Including a single extra parameter to model the unknown cut-off in Cepheid selection in SN Ia host galaxies lowers the recovered $H_0$ by $\Delta H_0=-1.4\,\kmsMpc$. Including further selection corrections for the SN Ia samples decreases the recovered $H_0$ by $\Delta H_0=-0.2\,\kmsMpc$, while using an equal-volume distance prior rather than one uniform in distance modulus reduces $H_0$ by $\Delta H_0=-1.0\,\kmsMpc$. When included together, these effects substantially reduce the inferred tension with CMB measurements.
\end{abstract}

% Write your keywords here

\maketitle

\section{Introduction}  \label{sec:intro}

Selection effects are a fundamental concern in observational astronomy because a detected sample of objects is rarely a representative subset of the underlying population \citep[see reviews by][]{Teerikorpi1997,AndreonHurne2013,Mantz2019}. Finite instrumental sensitivity, survey design, and detection thresholds preferentially select objects with particular properties, thereby introducing systematic biases into statistical analyses and population studies. In addition, the geometry of the Universe and the finite sensitivity of observations combine to produce distance-dependent selection effects. One of the most well-known examples is the Malmquist bias, whereby intrinsically luminous objects are overrepresented in flux-limited surveys because they remain detectable over larger volumes than fainter sources. Similarly, Eddington bias arises from measurement uncertainties that scatter objects across detection thresholds, distorting inferred distributions. Modern astronomical analyses therefore incorporate survey selection functions and completeness corrections whenever selection acts directly on the measured quantity. 

Within a Bayesian context, we can consider the posterior conditioned on the fact that the object was selected, allowing the selection to depend on both observed and latent variables:
\begin{equation}
    f(\theta\mid d,S=1)\propto
    \frac{f(d\mid\theta)f(S=1\mid d,\theta)}{f(d\mid S=1)}f(\theta)\,,
\end{equation} 
where $f(\theta\mid d,S=1)$ is the selection-conditioned posterior over model parameters $\theta$ given the data $d$ and that it passed selection $S=1$. $f(S=1\mid d,\theta)$ is the selection probability, the probability of selection given data and model parameters. The selection-conditioned evidence $f(d\mid S=1)$ ensures proper normalization \citep{Mandel2019}. This formulation remains valid even when selection is imposed on a different but correlated observable \citep{Mantz2019}. In general, selection effects alter the likelihood, while priors are placed on the underlying population unless the inference is about the selected sample in which case placing a prior on the observed population is appropriate. 

The traditional distance ladder approach combines geometric calibrators, Cepheid variable stars, and Type Ia supernovae to infer distances in the local Universe \citep[hereafter R22]{Riess2022}. The first rung of the ladder consists of geometric distances from trigonometric parallaxes, detached eclipsing binaries, maser systems, and other direct techniques \citep{GaiaCollaboration2021,Pietrzynski2019,Reid2019}. These measurements are used to calibrate the luminosities of Cepheid variable stars through the Leavitt period--luminosity relation \citep{Leavitt1912,Sandage2009,Freedman2012}. Cepheid distances to galaxies that have hosted Type Ia supernovae then provide a calibration of the supernova absolute magnitude \citep[][R22]{Freedman2001,Riess2016}, allowing Type Ia supernovae to serve as standardizable candles over cosmological distances \citep{Phillips1993,Tripp1998}. Combining calibrated supernova distances with measured redshifts over a patch of the Universe that is large enough to provide a representative level of expansion yields a percent-level determination of $H_0$ \citep[][R22]{Freedman2001,Riess2016}. Sample selection occurs at every step of the distance ladder, with selection typically made using observations that are not then used directly. For example, the selection of Cepheids within each SN Ia host is undertaken at visible wavelengths, while photometric measurements are made in the NIR (R22). The distance ladder analysis of R22 additionally selects the calibration sample of Milky Way Cepheids with parallax measurements, Cepheids in non SN Ia host galaxies, the SN Ia to follow-up with Cepheid observations \citep{Kenworthy2022}, and the SN Ia in the Hubble flow whose distances and redshifts are used to measure $H_0$. 

Independent measurements of $H_0$ come from the Cosmic Microwave Background (CMB), within the framework of the $\Lambda$CDM model \citep{Planck:2018vyg}. Additional techniques include strong gravitational lens time delays \citep{Wong2020}, and Baryon Acoustic Oscillation measurements combined with Big Bang Nucleosynthesis constraints on the baryon fraction to constrain the sound horizon \citep{DESI2025DR2}, and constraints using energy densities \citep{Crespi2026,Krolewski2026}. Comparisons among these methods have revealed the so-called ``Hubble tension,'' a persistent discrepancy between local and large-scale measurements of $H_0$, motivating extensive observational and theoretical investigations \citep{Verde2019,Freedman2021}. A recent compilation of local distance ladder results was provided by \citet{Casertano2026} who found $H_0=73.50\pm0.81\,\kmsMpc$, in $7.1\sigma$ tension with the  Planck+SPT+ACT CMB measurement of \citet{Camphuis2026}.

In this work we develop a simple statistical model illustrating how selection effects in correlated observations can bias measurements (Section~\ref{sec:selection}), and consider how this can be used to help to model selection effects in the distance ladder in Section~\ref{sec:distance-ladder-selection}. We then examine the prior assumptions adopted in the analysis (Section~\ref{sec:prior}), as these materially affect the results, and then look for residual selection effects in a modified version of the R22 distance ladder (Section~\ref{sec:distance-ladder}). Although the nuisance parameters correcting for residual selection effects may be only weakly constrained by the data, they are substantially degenerate with the inferred distance scale, so marginalizing over them can appreciably shift or broaden the posterior on $H_0$. Thus, key questions include not only whether residual selection effects are detected, but also how sensitive the inferred value of $H_0$ is to allowing for them. We present the answers in Section~\ref{sec:results1} \&~\ref{sec:results2} and conclude in Section~\ref{sec:concl}.

\section{Selection in correlated variables}
\label{sec:selection}

We now derive the key result of this work, a method to correct for selection effects where samples were selected using correlated observations. The derivation is presented in general terms to emphasize its applicability in many situations.

Let $x$ and $y$ be two astronomical measurements, with true values $\mu_x$ and $\mu_y$. Suppose that the observations are distributed as a multivariate Normal with covariance matrix characterized by variances $\sigma_x^2$ and $\sigma_y^2$ and correlation coefficient $\rho$. We consider a probit selection model where an observational constraint is placed on $x$, $x < x_{\rm lim}$, and we wish to know the distribution of $y$. Note that we set this problem up in such a way that it can be directly translated to magnitude measurements, where we consider an upper limit to $x$ equivalent to discarding faint measurements. The solution for a lower limit (applicable to working in flux) is also considered below. This analysis closely follows the reasoning introduced in econometrics by \citet{Heckman1979}.

The likelihood of interest in our analysis is given by $f(y\mid x<x_{\rm lim})$. We can derive this using Bayes' theorem
\begin{equation}
    f(y\mid x<x_{\rm lim})=\frac{f(x<x_{\rm lim} \mid y)f(y)}{f(x<x_{\rm lim})}\,.
\end{equation}
The conditional distribution of $x\mid y$ is Normal with mean $\mu_x + \rho \frac{\sigma_x}{\sigma_y}(y-\mu_y)$ and variance $\sigma_x^2(1-\rho^2)$. Integrating the conditional Normal distribution of $x\mid y$ up to the selection limit yields
\begin{equation}
  f(x<x_{\rm lim} \mid y) = \Phi\!\left(\frac{\alpha-\rho z}{\sqrt{1-\rho^2}}\right)\,,
\end{equation}
where $\alpha=(x_{\rm lim}-\mu_x)/\sigma_x$, $y=\mu_y+\sigma_yz$, and $\Phi$ is the standard Normal CDF. Using Bayes' theorem, the conditional distribution for $y$ is then 
\begin{equation}  \label{eq:y_pdf}
  f(y\mid x<x_{\rm lim}) =\frac{\phi(z)}
  {\sigma_y\Phi\!\left(\alpha\right)}
  \Phi\!\left(\frac{\alpha-\rho z}{\sqrt{1-\rho^2}}\right)\,.
\end{equation}
where $\phi$ is the standard Normal probability density function (PDF). This should be adopted for the likelihood of $y$ in a Bayesian analysis where correlated selection has occurred. The equivalent expression for a lower limit is
\begin{equation}
  f(y\mid x>x_{\rm lim}) =\frac{\phi(z)}
  {\sigma_y\left[1-\Phi\!\left(\alpha\right)\right]}
  \Phi\!\left(\frac{\rho z-\alpha}{\sqrt{1-\rho^2}}\right)\,.
\end{equation}

In terms of finding selection effects given just the Likelihood of $y$, things get quite nasty when there is a weak correlation $\rho\ll1$, and a regime where the selection removes only a small fraction of the population (i.e. $\alpha\gg1$). In this limit, the likelihood tends quickly to a Normal distribution with mean and variance
\begin{eqnarray}
  E\!\left[y \mid x < x_{\rm lim}\right]&=&
    \mu_y-\rho \sigma_y \lambda(\alpha) \label{eq:mag_offset}\\
  {\rm Var}\!\left(y \mid x < x_{\rm lim}\right)&=&
    \sigma_y^2\left[1-\rho^2\lambda(\alpha)\left(\lambda(\alpha)+        \alpha\right)\right]\,,
\end{eqnarray}
where $\lambda(\alpha)=\phi(\alpha)/\Phi(\alpha)$. For completeness, 
\begin{eqnarray}
  E\!\left[y \mid x > x_{\rm lim}\right]&=&
    \mu_y+\rho \sigma_y \lambda'(\alpha)\\
  {\rm Var}\!\left(y \mid x > x_{\rm lim}\right)&=&
    \sigma_y^2\left[1-\rho^2\lambda'(\alpha)\left(\lambda'(\alpha)-\alpha\right)\right]\,,
\end{eqnarray}
where $\lambda'(\alpha)=\phi(\alpha)/[1-\Phi(\alpha)]$

To first order the mean-median offset, defined as (mean$-$median)/$\sigma$, in the likelihood defined by Eq.~\ref{eq:y_pdf} is suppressed relative to the mean shift and arises only at higher order in $\rho$, whereas the mean offset is linear in $\rho$. Thus, observations where selection was made in a correlated variable can have a likelihood that is very close to a Normal distribution, hiding a selection effect, but still result in a significant offset in the mean.

\begin{center}
    \begin{figure}[!t]
        \centering
    	\includegraphics[scale=0.7]{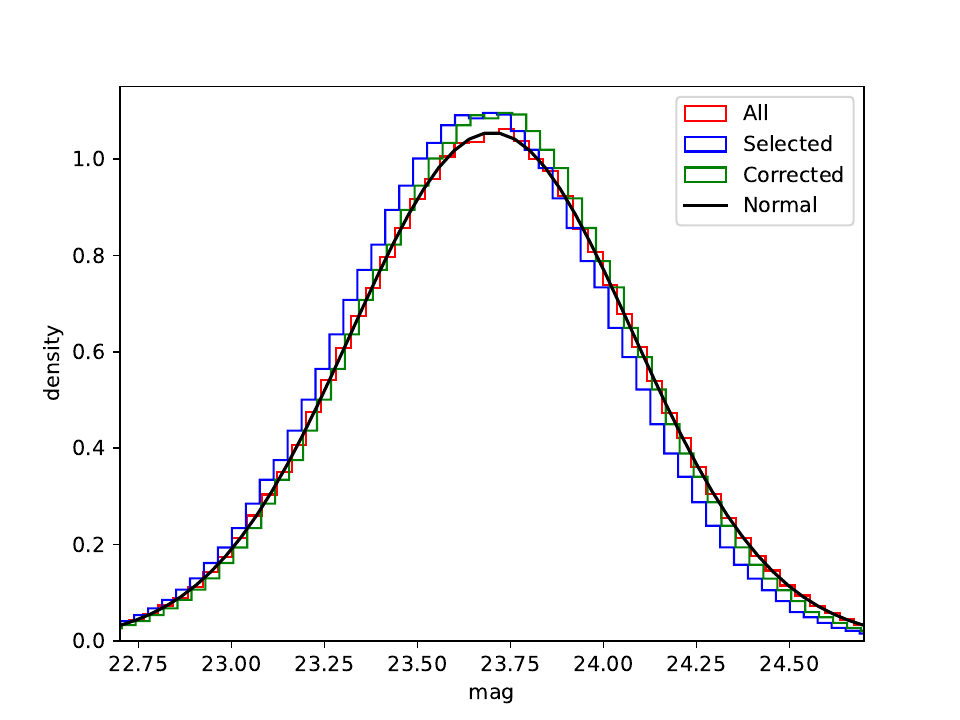}
    	\caption{The recovered distribution of measured magnitudes from a simple model of selection in a correlated, jointly Normal variable. In the absence of selection, the data follow the red “All” histogram, which is well described by a Normal distribution (black curve), as expected. After imposing selection, the observed distribution (blue histogram) is shifted toward brighter (lower magnitude) sources. In the weak-correlation regime, selection in a correlated variable primarily induces a shift in the mean, with only a small induced skewness. The green histogram shows the distribution after correcting for the selection-induced shift, such that the mean matches that of the full underlying sample.}
        \label{fig:MC-CV-Like}
    \end{figure}
\end{center}

\subsection{A Monte-Carlo demonstration}

In order to illustrate the impact of this effect, we demonstrate the above equations using a simple numerical model. We create $10^6$ draws from a bivariate Normal distribution with $\mu_y=23.7$, corresponding to the inverse-variance weighted mean IR magnitude of a Cepheid variable in a Type Ia supernova host galaxy in R22. A Normal distribution in magnitude corresponds to a lognormal distribution in flux, as expected for a dense stellar background and consistent with the forward-model simulations of R22. The variance $\sigma_y^2$ is set to match the weighted average variance for Cepheids in SN Ia host galaxies of $0.143$. 

Following Eq.~\ref{eq:y_pdf}, we assume that the expected dependence on the selection in $x$ is only through the dimensionless threshold $\alpha$, and we adopt $\alpha=1.2$, equivalent to requiring a very weak $1.2\sigma$ detection in $x$. The values of $\mu_x$ and $\sigma_x^2$ only enter through $\alpha$. The correlation coefficient appropriate for the Cepheid observations of R22 is not directly known; we adopt $\rho=0.5$ for illustration. These choices are intentionally selected to probe a regime where deviations from Normality remain small while selection-induced bias is appreciable. For $\alpha=1.2$, $\rho=0.5$, we expect the mean and median to be within $0.007\sigma$, while the predicted magnitude offset is $0.04$\,mag. The shift between mean and median is smaller than the level of $0.03\sigma$ reported in R22 based on the likelihood derived from forward models of Cepheid photometry without selection.

Fig.~\ref{fig:MC-CV-Like} shows the recovered distribution of magnitudes $y$, from (1) the full mock sample, (2) the sample after applying a selection cut in the correlated variable $x$, and (3) after correction by adding in the difference between true and expected mean magnitude. The measured median from the selected sample depends on the random seed, but is of order $0.007\sigma$ as expected. Applying the correction restores agreement between the selected and full distributions.

\section{Selection in distance ladder samples}  \label{sec:distance-ladder-selection}

The model given in Section~\ref{sec:selection} gives us a method to test for and correct for residual selection effects of a specific type, connected to truncated selection with a fixed correlation coefficient $\rho$ for each sample. This is an effective statistical description, not a physical model for any sample within the SH0ES pipeline. But, crucially, for those samples where each object has a different photometric uncertainty, the predicted selection correction will differ from object to object. The heteroscedastic uncertainties therefore provide information that helps distinguish selection corrections from changes in the underlying distance scale. If residual selection effects are present of the form described in Section~\ref{sec:selection}, then including a selection-like correction is expected to improve the global fit to the distance ladder. Because the correction is constrained to scale with the independently measured photometric uncertainties, it cannot absorb arbitrary residuals. Missed complications in the selection are likely to degrade the effect of the correction, but it is still worthwhile to apply the simple model to investigate the evidence and impact of possible effects where no other selection information is easy to obtain.

For selection corrections where there are free parameters, we infer the selection parameters simultaneously with the distance-ladder parameters from the same data. This is the standard Bayesian treatment of nuisance parameters: rather than estimating and correcting for selection effects in a separate step, uncertainty in the selection model is propagated directly into the posterior for $H_0$. Consequently, evidence for non-zero selection corrections and the inferred value of $H_0$ are obtained from a single joint posterior. For the model of Section~\ref{sec:selection}, it is possible that a different systematic effect could provide a similar correlation with the uncertainties, for Cepheids due to crowding, background subtraction, period dependence through signal-to-noise ratio, or galaxy surface brightness, for example. In this sense, applying the result of Section~\ref{sec:selection} simply shows that there is a plausible route to a systematic offset that scales with the error. This is worth exploring in future work through full simulations of selection and magnitude measurement. We limit ourselves to using this model to consider if there is evidence in the data itself for a systematic scaling, and if so, what would be the potential impact on $H_0$.

The distance ladder approach to measure $H_0$ requires the selection of a number of samples of  astronomical objects. To model the selection of some of these, we apply the methodology of Section~\ref{sec:selection}. For others, where the selection can be more easily modelled, we include that selection explicitly or make use of previous results that do so. We now list and consider each sample separately.

\subsection{Milky Way Cepheids} \label{sec:MWselection}

R22 used Gaia \citep{GaiaCollaboration2021} parallax measurements coupled with high precision photometry of MW Cepheids to calculate the zero-point of the Leavitt relation. This zero-point is defined using colour-corrected Wesenheit magnitudes, $M_{H,i}^{W,\text{obs}}$. This needs to be adjusted for the period and metallicity of a specific Cepheid $i$, such that the model for the observed Cepheid brightness is
\begin{equation}
  m_{H,i}^W = \mu_i + M_{H}^W + b_W (\log_{10}P_i -1)
    + Z_W [O/H]_i\,,
\end{equation}
where $b_W$ and $Z_W$ are model parameters controlling the period-luminosity relation and metallicity correction respectively. $\mu_i$ is the distance modulus, $P_i$ the period, and $[O/H]_i$ the metallicity. The distance moduli to all Cepheids are jointly fitted using models for both the brightness and parallax measurements. The model for the observed parallax $\pi_{\text{EDR3},i}$ is
\begin{equation}
  \pi_i= 10^{(10 - \mu_i)/5} - zp\,,
\end{equation}
where the model parameter $zp$ allows for potential correlated systematic effects in the parallax measurements. 

\citet{Stiskalek2026} recently extended this analysis to include the selection of MW Cepheids in the likelihood, and a prior designed to match the expected stellar distribution within the MW (see Section~\ref{sec:prior} for a further discussion of this). They recovered $M_H^W=-5.909\pm0.022$ and $Z_W=-0.22\pm0.08$, which we adopt for our baseline results and denote ``S26''.

\subsection{Local SN Ia host galaxies} \label{sec:sel-local-sn}

The SH$0$ES team linked SN Ia and Cepheid brightnesses by observing both within common host galaxies. Hosts were selected because of the presence of a SN Ia, potentially introducing selection effects within the SN Ia sample \citep{Kenworthy2022,Stiskalek2025}. Because the host galaxies are simply used to match Cepheid and SN Ia calibrations, the host galaxy distribution in redshift (or distance) does not directly bias the measurement of $H_0$. However, selection based on SN Ia brightness will potentially be an issue as we need the sample of SN Ia to match that of the Hubble flow SN Ia used to measure $H_0$. \citet{Stiskalek2025} model the selection as a simple cut in Wesenheit magnitude of $m_{H,{\rm lim}}^W$, with the results not sensitive to the exact value. Two of the most distant galaxies, N0105 and N0976, need to be removed from the analysis as they were targeted by a different program with a very different selection method \citep{Kenworthy2022,Stiskalek2025}. We remove these galaxies, and adjust the likelihood by including a normalization term in the likelihood, multiplying the likelihood by
\begin{equation}
    1/\Phi\left(\frac{m_{H,{\rm lim}}^W-\hat{m}_{H}^W)}
        {\sqrt{\sigma_m^2+\sigma_{\rm sel}^2}}\right)\,,
\end{equation}
where $\Phi$ is the standard multivariate Normal cumulative distribution function (CDF) with a correlation matrix matching that of the SN Ia Wesenheit magnitude measurements, acting upon the vector of normalised differences between the magnitude limit and the model magnitude $\hat{m}_{H}^W$. The selection dispersion ($\sigma_{\rm sel}=0.15$\,mag) allows for a degree of randomness in the selection \citep{Stiskalek2025} rather than a sharp cut-off. We find that the correction is not sensitive to the exact values of either $\sigma_{\rm sel}$ or $m_{H,{\rm lim}}^W$, and we adopt $m_{H,{\rm lim}}^W=14.15$\,mag, corresponding to the maximum SN Ia magnitude in the sample. We solve this multi-variate integral numerically using a Cholesky decomposition with Genz-Bretz pivoting of the correlation matrix.

\subsection{Cepheids within SN Ia hosts} 

We include data from 35 SN Ia host galaxies for which there are also Cepheid observations. For the 19 SN Ia of these reported in \citet{Hoffmann2016}, hosts were imaged 11–12 times over 60–100 days. First, variable stars were isolated, and then fitted using templates to identify Cepheids, and various selection criteria were placed on the objects to ensure they match the expected properties of isolated Cepheids with low to moderate reddening \citep[For details, see][]{Hoffmann2016}. Incompleteness may arise from inadequate phase coverage at low periods, so a minimum period is imposed, calculated based on expected signal-to-noise. R22 collated these data and additional SN Ia host galaxies, employing an automated procedure based on the same methodology for all. Outlier rejection set at $3.3\,\sigma$ was then applied to remove extrema beyond the Leavitt relation. Independent reviews of the Cepheid selection and photometric measurement procedure can be found in \citet{Efstathiou2014,Efstathiou2020}, that discuss further selection criteria and their possible impact. In total, R22 include 2150 Cepheid magnitudes measured in galaxies hosting SN Ia.

Once Cepheids were identified, photometric measurements in the NIR were made: observations in the NIR reduce the systematic uncertainty associated with the reddening laws for Cepheids and their hosts and the Cepheid metallicity dependence. While recent JWST observations support the HST photometric calibration \citep{Riess2023,Riess2024,Riess2025}, they do not reselect the Cepheids and any effects arising from that will remain. The sequence of observations - selection in the visible followed by separate measurements in the NIR - means that we have selection and observation correlated through the background crowding, matching the derivation in Section~\ref{sec:selection}. Although the R22 cut on low period Cepheids is designed to remove selection effects, given the importance of these observations, it is worth testing for residual effects.

\subsection{Cepheids in other galaxies} 

In the analysis of R22, samples of Cepheids were included in NGC 4258, M31, the LMC (ground based and using HST), and the SMC. For the selection of Cepheids in NGC 4258, see \citet{Yuan2022}: we consider this selection in the same way as for the Cepheid selection in the SN Ia hosts. For the others, we cannot justify that the selection is the same in them all. In total, R22 included 980 Cepheid magnitudes in non SN Ia host galaxies.

\subsection{SN Ia within the Hubble flow} 

The final sample used in the R22 analysis included 277 SN Ia magnitudes covering $0.0233<z<0.15$ selected from the Pantheon+ sample \citep{Scolnic2022} in the Hubble flow. The selection of this sample is described in section 6.8 of R22 and consists of a redshift selection $0.0233<z<0.15$ for objects that pass the Pantheon+ quality cuts. Constraints are then placed on the host galaxies in order to match their distribution to that of the Cepheid hosts. Given the complicated selection, we have decided to simply test for residual truncation-like selection effects using the methodology of Section~\ref{sec:selection}.

\section{Priors}  \label{sec:prior}

In order to make Bayesian inferences from the posterior, we need to define priors on the model parameters. The key priors for the distance-ladder are those on distances, and R22 adopted priors on distances that were uniform in distance modulus for both the analysis combining parallax and brightness measurements of Milky Way (MW) Cepheids and the analysis combining SN Ia and Cepheid brightnesses. Distance moduli were the parameters used to form the sample space of the recovered posterior, so adopting a uniform prior in these parameters means that no additional terms need to be added to the likelihood to form the posterior. Note that priors transform as PDFs: if $y=g(x)$, we have the prior in $y$:
\begin{equation}  \label{eq:prior-transform}
f_Y(y) = f_X\big(g^{-1}(y)\big) \left| \frac{dx}{dy} \right|\,.
\end{equation}
It is immediately apparent that a uniform prior in one variable is not necessarily uniform in another. 

As discussed in Section~\ref{sec:MWselection}, for MW Cepheids, we contrast results using the constant distance modulus prior against the results from \citet{Stiskalek2026}, who reanalysed the MW Cepheids using a prior designed to match the expected stellar distribution within the MW. To see how such a prior can be constructed, \citet{BailerJonesetal2021} advocated for a Generalized Gamma Distribution prior for the MW, which is an exponentially cut-off power-law in distance $r$. The parameters are calculated as a function of angular position in order to match a model for the density of stars
\begin{equation}  \label{eq:prior}
  f(r) = \begin{cases} 
    \frac{1}{\Gamma\left(\frac{b+1}{a}\right)} \frac{a}{L^{b+1}} r^b e^{-(r/L)^a} & \text{if } r \geq 0 \\ 0 & \text{otherwise} \end{cases}
\end{equation}
where $a$, $b$ and $L$ are parameters whose values are determined from a Galactic model and depend on the angular position. This is a distribution in distance $r$, so if applying to a MCMC run that steps through the model distance modulus, we would need to multiply by a further factor of $|dr/d\mu|=|0.2\ln(10)r|$ to account for the derivative in Eq.~\ref{eq:prior-transform}. Near the observer, $b=2$ corresponds to a constant spatial number density, while other values permit steeper or shallower radial profiles. The sharpness of the cut-off is controlled by $a$, while $L$ controls the expected distance. Values of these parameters are tabulated on a HEALPix grid \citep{Gorski2005HEALPix}.

For the Cepheid-host SN Ia and Hubble-flow SN Ia, we consider a simplified population prior with weight given to equal-volumes (constant in density) as advocated by \citet{Desmond2025DistancePrior}. Considering scales where evolution is unimportant, this prior matches the idea of there being no preferred location in a Euclidean universe for which $f(r)\propto r^2$. Equivalently, one may regard this as sampling in $r^3$ and adopt a uniform prior in that variable. Given that the exploration of the posterior is usually set up as an exploration in distance modulus, the explicit prior to add in this case is $\ln f(\mu)=0.6\ln(10)\mu+{\rm const}$, where $\mu$ is the distance modulus. We contrast using this prior for extragalactic distances and one uniform in $\mu$: for simplicity we refer to them as uniform-$r^3$ and uniform-$\mu$ priors.

\section{Revised recovery of $H_0$ using the distance ladder} \label{sec:distance-ladder}

\subsection{A revised Bayesian analysis for $H_0$} \label{sec:DL-fits}

We follow the convention of R22 and perform a combined linear fit to the extragalactic Cepheid and SN Ia data to infer $H_0$. The data vector, ${\bf y}$ consists of 2150 Cepheid magnitudes measured in galaxies hosting SN Ia, 980 Cepheid magnitudes in other galaxies, 77 SN Ia magnitudes in nearby galaxies, and 277 SN Ia magnitudes covering $0.0233<z<0.15$ from the Pantheon+ sample in the Hubble flow. R22 also include external constraints on $M_H^W$ derived from HST and ground-based parallaxes, a constraint on $Z_W$, and anchor distance measurements for NGC 4258 and the LMC. In the following we consider replacing the R22 constraints on $M_H^W$ and $Z_W$ with those obtained by \citet{Stiskalek2026}. If we use the R22 matrices as provided, we match their measurement of $H_0=73.04\pm1.01\,\kmsMpc$ (see Table~\ref{table:H0-results1}).

The standard fit to these data uses a model with 46 parameters ${\bf q}$\,=\,($\mu_i$, $M_H^W$, $M_B$, $b_W$, $Z_W$, $\Delta zp$, $5\log H_0$). Forty of these are distance moduli corresponding to the 37 SN Ia host galaxies together with NGC 4258, M31 and the SMC and LMC distance. When including the selection effect correction for the SN Ia in Cepheid host galaxies, the number of SN Ia host galaxies is reduced to 35 (see Section~\ref{sec:sel-local-sn}). The remaining six parameters constrain the Cepheid period-luminosity relation (PLR) intercept and slopes $M_H^W$, $b_W$, $Z_W$, the difference between the HST and ground-based magnitude zero points $\Delta zp$, the fiducial SN Ia absolute magnitude $M_B$, and finally the Hubble constant, via the parameter $5\log H_0$. No constraint on the slope $b_W$ is taken from the Milky Way Cepheid analysis, allowing for possible differences in the period–luminosity slope between the Milky Way and extragalactic Cepheid samples. More details of this approach can be found in R22. The model for the observed magnitudes is linear in these parameters, and can be written $L{\bf q}$, where $L$ is an $n_{\rm data}\times n_{\rm model}$ matrix. 

To allow for residual selection effects of the form discussed in Section~\ref{sec:selection}, we introduce extra model parameters that correspond to the combination $\rho\lambda(\alpha)$ in Eq.~\ref{eq:mag_offset}. For the Cepheids in SN Ia host galaxies and NGC 4258, we contrast three alternative parameterizations: (i) independent parameters $s_i$ for each host $i$ with uniform priors; (ii) a hierarchical model in which the $s_i$ are drawn from a Normal distribution with uniform hyperpriors on the unknown mean and variance; and (iii) a single shared parameter $s$ for all hosts with a uniform prior. For each Cepheid $j$ in the SN Ia host galaxies and for NGC 4258 with magnitude uncertainty $\sigma_j$, we add a correction to the Wesenheit magnitude model of $s_i\sigma_j$ for (i) or (ii) and $s\sigma_j$ for (iii). For the Hubble flow SN Ia, we only consider a model with a single extra parameter $s_{\rm HF}$  for all SN Ia.

For the Cepheid selection model with multiple parameters $s_i$, we can extend the standard formalism by defining $U$, a $n_{\rm data}\times n_{\rm host}$ matrix whose entries are zero except in rows corresponding to affected magnitudes. For Cepheid $j$ in host galaxy $i$, the corresponding row contains the value $\sigma_{j}$. Now we can extend our model calculation by concatenating $L$ and $U$ to give $L'=[L U]$, an $n_{\rm data}\times(n_{\rm model, orig}+n_{\rm host})$ matrix, and extending our model parameter vector to include the $s_i$. The likelihood is then
\begin{equation}
  \ln\mathcal{L} \propto -\frac{1}{2}({\bf y} - L'{\bf q})^TC^{-1}({\bf y} - L'{\bf q})\,.
\end{equation}
Since the uniform-$r^3$ priors contribute linearly in the distance moduli, the logarithm of the prior can be written as $\ln f({\bf q})\propto {\bf p}^T{\bf q}+{\rm const}$, where {\bf p} has entries $0.6\ln(10)$, corresponding to the Jacobian of a uniform-volume prior, for the distance-modulus parameters and zeros elsewhere. For the uniform-$\mu$ prior, we can set ${\bf p}={\bf 0}$. For the selection correction model with a single parameter $s$, the matrix $L$ is instead concatenated with a single column vector whose entries are zero except for the values $\sigma_{j}$ corresponding to magnitude observations in the affected objects.

For all such models except that with the hierarchical prior on the $s_i$ and the selection correction for the SN Ia in Cepheid host galaxies, the posterior remains multivariate Normal because both the likelihood and the log-prior are linear or quadratic in the parameters. Consequently, parameter constraints and marginal covariances can be obtained analytically. The posterior mean (equivalently the maximum a posteriori, MAP, estimate) is:
\begin{equation}  \label{eq:best-fit}
  \hat{\bf q}=(L'^T C^{-1}L')^{-1}(L'^TC^{-1}{\bf y}+{\bf p})\,.
\end{equation}
Since the adopted prior contributes only a linear term, it does not alter the Hessian of the posterior and the covariance matrix of $\hat{\bf q}$ is $(L'^T C^{-1}L')^{-1}$. We can take the root of the diagonal terms to obtain marginalized standard deviations. In Section~\ref{sec:results1} we compare the analytic solution with MCMC sampling and find excellent agreement. In practice, we find that the augmented design matrix remains well conditioned and the analytic solution is numerically stable. The analytic approach is several orders of magnitude faster than MCMC sampling and therefore enables a much broader suite of consistency checks and sensitivity tests (see Section~\ref{sec:PPT}). 

We consider an alternative two-step procedure in Appendix~\ref{app:2-step}, which  first estimates the selection corrections and then fits the distance ladder using corrected magnitudes. However, this does not correctly propagate uncertainty in the selection corrections into the posterior for $H_0$, in contrast to our default method of a joint Bayesian analysis. As discussed in Appendix~\ref{app:2-step}, the two-step approach artificially removes the covariance between these quantities and therefore does not correspond to inference under a single generative model.

\subsection{Results including Cepheid selection}  \label{sec:results1}

\begingroup 
    \setlength{\tabcolsep}{10pt} % Default value: 6pt
    \renewcommand{\arraystretch}{1.5} % Default value: 1
    \setlength\extrarowheight{2pt}
    \begin{table}
        \centering
        \begin{tabular}{ c c c c }  
            prior & selection & \multicolumn{2}{c}{$H_0\,\kmsMpc$}\\
            & model & MCMC & analytic\\
            \hline \hline
            Uniform-$\mu$, $\log H_0$, R22 MW & none & $H_0=73.05\pm1.01$ & $H_0=73.04\pm1.01$ \\ % run 00100000
            Uniform-$r^3$, $\log H_0$, R22 MW & none & $H_0=72.04\pm0.99$ & $H_0=72.04\pm0.99$ \\ % run 10100000
            Uniform-$r^3$, $\log H_0$, S26 MW & none & $H_0=71.73\pm1.05$ & $H_0=71.73\pm1.04$ \\ % run 10200000
            Uniform-$r^3$, $H_0$, S26 MW & none & $H_0=71.73\pm1.04$ & $H_0=71.74\pm1.04$ \\ % run 11200000
            \hline
            Uniform-$\mu$, $\log H_0$, R22 MW & CV individual & $H_0=72.24\pm1.33$ & $H_0=72.21\pm1.33$ \\ % run 00110000
            Uniform-$\mu$, $\log H_0$, R22 MW & CV hierarchical & $H_0=72.24\pm1.37$ & - \\ % run 00120000
            Uniform-$\mu$, $\log H_0$, R22 MW & CV shared & $H_0=72.20\pm1.19$ & $H_0=72.20\pm1.17$ \\ % run 00130000
            Uniform-$\mu$, $\log H_0$, R22 MW & CV individual (42) & $H_0=71.64\pm1.42$ & $H_0=71.61\pm1.43$ \\ % run 00150000
            Uniform-$\mu$, $\log H_0$, R22 MW & CV shared (42) & $H_0=72.28\pm1.15$ & $H_0=72.27\pm1.14$ \\ % run 00160000
            \hline
            Uniform-$r^3$, $\log H_0$, S26 MW & CV individual & $H_0=69.73\pm1.34$ & $H_0=69.72\pm1.34$ \\ % run 10210000
            Uniform-$r^3$, $\log H_0$, S26 MW & CV hierarchical & $H_0=70.18\pm1.40$ & - \\ % run 10220000
            Uniform-$r^3$, $\log H_0$, S26 MW & CV shared & $H_0=70.46\pm1.17$ & $H_0=70.43\pm1.17$ \\ % run 10230000
            Uniform-$r^3$, $\log H_0$, S26 MW & CV individual (42) & $H_0=67.56\pm1.54$ & $H_0=67.55\pm1.53$ \\ % run 10250000
            Uniform-$r^3$, $\log H_0$, S26 MW & CV shared (42) & $H_0=70.66\pm1.14$ & $H_0=70.65\pm1.14$\\ % run 10260000
            \hline \hline  
        \end{tabular}  
        \caption{Recovered values of the Hubble constant under different assumptions for priors and treatment of extragalactic Cepheid selection effects. We consider two distance priors: uniform-$\mu$, or uniform-$r^3$, and three ways of adding in extra corrections for extragalactic Cepheid selection with no further selection corrections: {\it individual}, including parameters for each galaxy with a uniform prior on each; {\it hierarchical}, parameters for each galaxy with a constraining Normal distribution with uniform hyperpriors on the mean and variance; {\it shared}, a single selection coefficient applied to all galaxies, with a uniform prior. $(42)$ indicates that selection corrections were included for all $42$ galaxies including M31, SMC and LMC separately for both ground and space observations. ``analytic'' is unavailable for the hierarchical model because the posterior is no longer multivariate Normal. The first row of the table shows the fits using the R22 analysis method. The label "S26 MW" indicates the values of $M_H^W$ and $Z_W$ were taken from the MW analysis of \citet{Stiskalek2026}, which included corrections for MW Cepheid selection.}
        \label{table:H0-results1}
    \end{table}
\endgroup

Table~\ref{table:H0-results1} summarizes the inferred values of $H_0$ for the different priors and CV selection models. Analytic and MCMC solutions are in excellent agreement, validating the analytic treatment presented in Section~\ref{sec:distance-ladder}. The first tier of results creates a baseline without any corrections for selection effects, with the first line in Table~\ref{table:H0-results1} matching the R22 results. Adopting a uniform-$r^3$ prior reduces $H_0$ by $-1.0\,\kmsMpc$ compared to a uniform-$\mu$ prior, consistent with the results found by \citet{Desmond2025DistancePrior,hogas2026physically}. Adopting the analysis of \citet{Stiskalek2026} for the MW Cepheids reduces this difference by a further $-0.3\,\kmsMpc$. This is a small change because of the balance between  the competing effects of the selection corrections and the prior change on the MW Cepheid analysis. Changing the uniform prior on $\log H_0$ to one uniform in $H_0$ makes no difference to the results.

The second and third tiers of results in Table~\ref{table:H0-results1} present results when selection-correction parameters are used to correct Cepheid magnitudes for selection within their host galaxies. For individual selection corrections $s_i$ for each SN Ia host and NGC 4258, we see a reduction $H_0$ of $-0.8\,\kmsMpc$ and $-2.0\,\kmsMpc$ for a uniform-$\mu$ and uniform-$r^3$ prior respectively. While the shift is similar for the uniform-$\mu$ distance prior with a single parameter $s$ or a hierarchical prior, we see a large difference for the uniform-$r^3$ distance prior, where the shift is reduced to $-1.3\,\kmsMpc$ when changing to a single parameter $s$. The combination of the uniform-$r^3$ prior and the inclusion of selection corrections leads to a larger shift compared to summing the shifts from each change individually because the selection correction allows the distance prior to have more influence on the result. Including selection effect corrections for the Cepheids within the auxilliary galaxies, SMC, LMC and M31 has a far larger effect on $H_0$, reflecting an increased coupling between the distance prior and the selection model: because the selection correction systematically alters the inferred distance moduli, it changes the leverage of the distance prior on the global fit. The size of this shift allowed for individual corrections is reduced if we place a hyperprior on the selection correction terms, or if we limit $s_i$ to be the same for all galaxies $s$ because there is less freedom for this coupling. 

A shared selection parameter implicitly assumes that all host galaxies require the same additional selection correction in addition to those applied by R22. The independent-$s_i$ model is the more flexible parameterization that imposes the fewest assumptions about the similarity of the Cepheid selection functions within different host galaxies. However this couples with the uniform-$r^3$ prior making the result strongly prior dependent and, as we will see in Section~\ref{sec:significance}, the change in $\chi^2$ indicates that the model overfits the data. Adopting the MW based zero-point of \citet{Stiskalek2026} and the Cepheid selection model with a single parameter $s$ for the SN Ia hosts and NGC 4258 forms our conservative baseline.

\begin{center}
    \begin{figure}[!t]
        \centering
    	\includegraphics[scale=0.7]{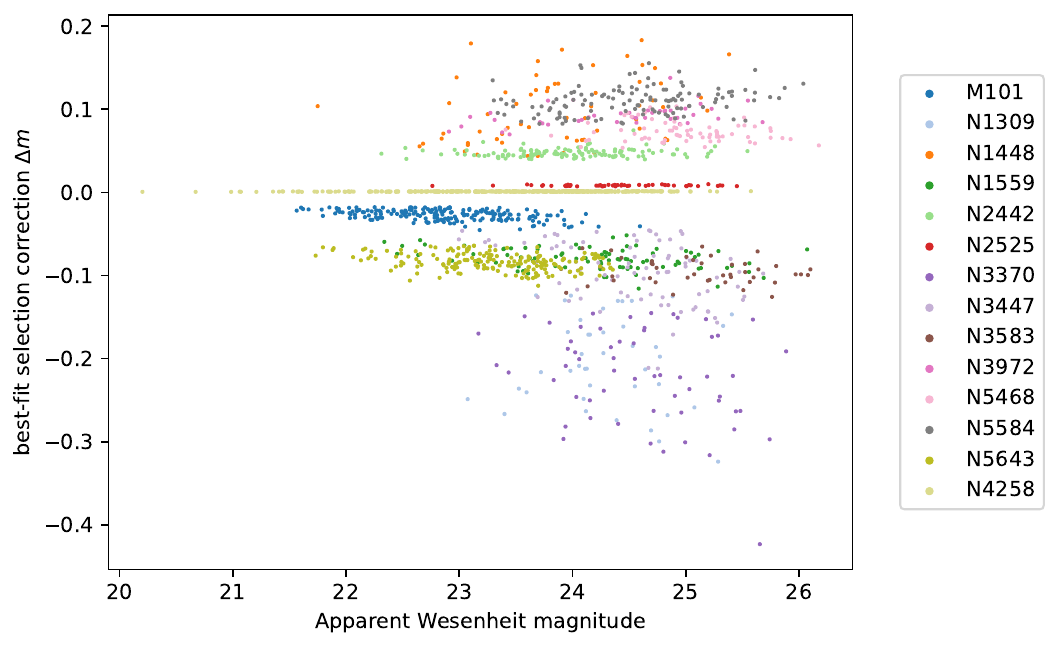}
    	\caption{The derived magnitude correction applied to Cepheid magnitudes in SN Ia host galaxies to account for selection effects. Data are plotted for galaxies hosting more than 40 observed Cepheids, colour-coded by each galaxy. The corrections are inferred from the data, as described in Section~\ref{sec:DL-fits} allowing a different selection coefficient $s_i$ for each host. Both positive and negative corrections are observed, an expected consequence of statistical fluctuations in the data. The net effect of these corrections on the inferred value of $H_0$ is described in Section~\ref{sec:results1}. \vspace{2mm}}
        \label{fig:Delta_m}
    \end{figure}
\end{center}

\subsection{Significance of results for Cepheid selection}  \label{sec:significance}

The inferred distribution of selection corrections for SN Ia host galaxies with more than 40 observed Cepheids is shown in Fig.~\ref{fig:Delta_m}. Allowing individual corrections for each SN Ia host galaxy and NGC 4258, the inferred correction is (+0.03) mag for the uniform-$\mu$ prior, and (+0.05) mag if a uniform-$r^3$ prior is used instead. Although both positive and negative corrections are recovered for individual hosts, the mean correction is positive, corresponding to systematically fainter inferred Cepheid magnitudes and hence a lower value of $H_0$. However, the addition of selection effect corrections of this form, by themselves, do not shift magnitudes sufficiently to solve the Hubble tension.

We define the detection significance by considering the weighted average of the $s_i$, where all $s_i$ are expected to have the same value. The signal-to-noise is then $Z={\bf 1}^TC_s^{-1}\hat{\bf s}/\sqrt{{\bf 1}^TC_s^{-1}{\bf 1}}$, where ${\bf 1}$ is a vector of 1's of length matching the number of selection effect parameters, ${\bf s}$ is the vector of selection effect parameters, and $C_s$ is the component of the model parameter covariance for the selection effect parameters. Because we are recovering the significance from fitting an overall amplitude, the same significance is determined when fitting a single $s$ or multiple $s_i$, one for each host. For the fit with correction for the $38$ distant host galaxies, we find that the posterior gives a $1.3\sigma$ preference for a non-zero correction for the uniform-$\mu$ prior, $2.3\sigma$ for a uniform-$r^3$ prior, independent of the MW Cepheid zero-point adopted. 

Including corrections for the auxilliary hosts (the ``(42)'' model in table~\ref{table:H0-results1}), only changes the significances very slightly from our default model to $1.3\sigma$ for the uniform-$\mu$ prior, and $2.2\sigma$ for a uniform-$r^3$ prior. Given this, and that the Cepheid selection was very different in these hosts, we adopt only applying Cepheid selection corrections for the distant hosts as the baseline in our analysis. 

\subsection{Results with multiple selection corrections}  \label{sec:results2}

\begingroup 
    \setlength{\tabcolsep}{10pt} % Default value: 6pt
    \renewcommand{\arraystretch}{1.5} % Default value: 1
    \setlength\extrarowheight{2pt}
    \begin{table}
        \centering
        \begin{tabular}{ c c c }  
            selection & \multicolumn{2}{c}{$H_0\,\kmsMpc$} \\
            model & uniform-$\mu$ & uniform-$r^3$ \\
            \hline \hline
            R22 & $H_0=73.04\pm1.01$ & $H_0=72.04\pm0.99$ \\ % run 00100000, 10100000
            S26 & $H_0=72.92\pm1.06$ & $H_0=71.73\pm1.04$ \\ % run 00200000, 10200000
            R22, CV shared & $H_0=72.20\pm1.17$ & $H_0=70.60\pm1.15$ \\ % run 00130000, 10130000
            R22, SN Ia host & $H_0=73.21\pm1.00$ & $H_0=72.26\pm1.00$ \\ % run 00101000, 10101000
            R22, SN Ia HF & $H_0=72.57\pm1.65$ & $H_0=71.61\pm1.63$ \\ % run 00100100, 10100100
            \hline
            S26, CV shared, SN Ia host, SN Ia HF & $H_0=71.80\pm1.77$ & $H_0=70.23\pm1.72$ \\ % run 00231100, 10231100
            S26, CV shared, SN Ia host, SN Ia HF, no auxilliary & $H_0=72.81\pm2.06$ & $H_0=69.76\pm1.96$ \\ % run 00233100, 10233100
            \hline \hline  
        \end{tabular}  
        \caption{Recovered values of the Hubble constant under different corrections for selection effects. All rows in this table use a uniform prior on $\log H_0$. Given the consistency between analytic and numerical results as demonstrated by Table~\ref{table:H0-results1}, we only present a single set of constraints from each posterior choice - from the analytic solution if available, otherwise the marginal solution averaged over the MCMC chains. Selection corrections are included by adopting the MW Cepheid zero-point of \citet{Stiskalek2026} (``S26'' in Table~\ref{table:H0-results1}), or excluded by adopting those from R22. Note that S26 also adopted different distance priors for the MW analysis: here ``Uniform-$r^3$'' or ``uniform-$\mu$'' refer only to the extragalactic distance priors. We next present results including a selection effect correction for a single sample: the Cepheid selection with a shared parameter $s$ (``CV shared''), the correction for SN Ia selection for Cepheid hosts (``SN Ia host'') or the selection of Hubble flow SN Ia (``SN Ia HF'') as discussed in Section~\ref{sec:distance-ladder-selection}. The final two entries in the table include all selection effect corrections discussed in that Section. The only remaining potential selection effects not addressed are the auxilliary hosts, M31, the LMC and SMC. The penultimate entry includes these data, while the final entry in the table excludes them.}
        \label{table:H0-results2}
    \end{table}
\endgroup

We now consider the inclusion of corrections for selection effects for the other components of the distance ladder, as described in Section~\ref{sec:distance-ladder-selection}. The first five entries in Table~\ref{table:H0-results2} present results where we consider the inclusion of selection effects individually. In general, selection excludes fainter, more distant objects, and so correcting for them in any sample pushes the MAP brightness to fainter values. When translated to a change in $H_0$ for the distance ladder approach, adding a correction for a selection effect will tend to increase $H_0$ for calibration samples (for example, samples used to set a zero-point), and decrease $H_0$ for samples used to measure distances.

Changing from the R22 MW zero-point to that of \citet{Stiskalek2026} only makes a very small change in $H_0$ as the shift from a uniform-$\mu$ to a uniform-$r^3$ prior they adopted for MW Cepheids was balanced by the inclusion of selection corrections. Introducing the extra correction of Section~\ref{sec:selection} for the extragalactic Cepheids decreases $H_0$ as discussed in Section~\ref{sec:results1}. Introducing the correction for the selection of SN Ia host galaxies increases $H_0$ by $+0.2\,\kmsMpc$, while applying the model of Section~\ref{sec:selection} to the Hubble flow SN Ia decreases $H_0$ by $-0.45\,\kmsMpc$ and increases the error significantly. The significance for this term is very weak compared to that for the Cepheids in SN Ia hosts at $0.4\sigma$. Given this, the inclusion of the terms is effectively a robustness test, rather than evidence for a Hubble flow SN Ia selection correction. We see no sensitivity of the change due to the selection of the Hubble flow SN Ia to the distance prior used, in contrast to the Cepheid selection. 

The final two rows in Table~\ref{table:H0-results2} give the $H_0$ measurements including all of the selection effect corrections discussed in Section~\ref{sec:distance-ladder-selection}. The only sample used in R22 that we have not included a selection effect correction for are the Cepheid observations in auxilliary hosts (M31, LMC, SMC), due to the complicated selection of Cepheids within these objects. Table~\ref{table:H0-results2} shows results with and without these data. We see a significant change in $H_0$ when doing this, with the exclusion pushing the results in line with the distance prior preference: increasing $H_0$ for the uniform-$\mu$ prior, and decreasing it slightly for the uniform-$r^3$ prior. As expected with the removal of data, the error is increased.

\subsection{Posterior Predictive Tests}  \label{sec:PPT}

Posterior predictive tests (PPTs) provide a method for assessing whether the assumptions underlying a Bayesian model are consistent with the observed data. The procedure involves drawing parameter values from the posterior distribution (e.g. by selecting samples from an MCMC chain) and treating each draw as a data-generating model. Synthetic data sets are then generated from the likelihood conditioned on these parameters, enabling comparison between simulated and observed data via chosen summary statistics. We consider four summary statistics: the distributions of Cepheid and SN Ia magnitudes, the goodness-of-fit statistic $\chi^2$, the difference in the recovered value of $H_0$ with and without selection corrections, and the amplitude of the extragalactic Cepheid selection terms in the model as a function of $\sigma_j$. Posterior predictive agreement is a necessary condition for a good model but not evidence that the model is uniquely correct.

\begin{center}
    \begin{figure}[!t]
        \centering

        \includegraphics[scale=0.43]{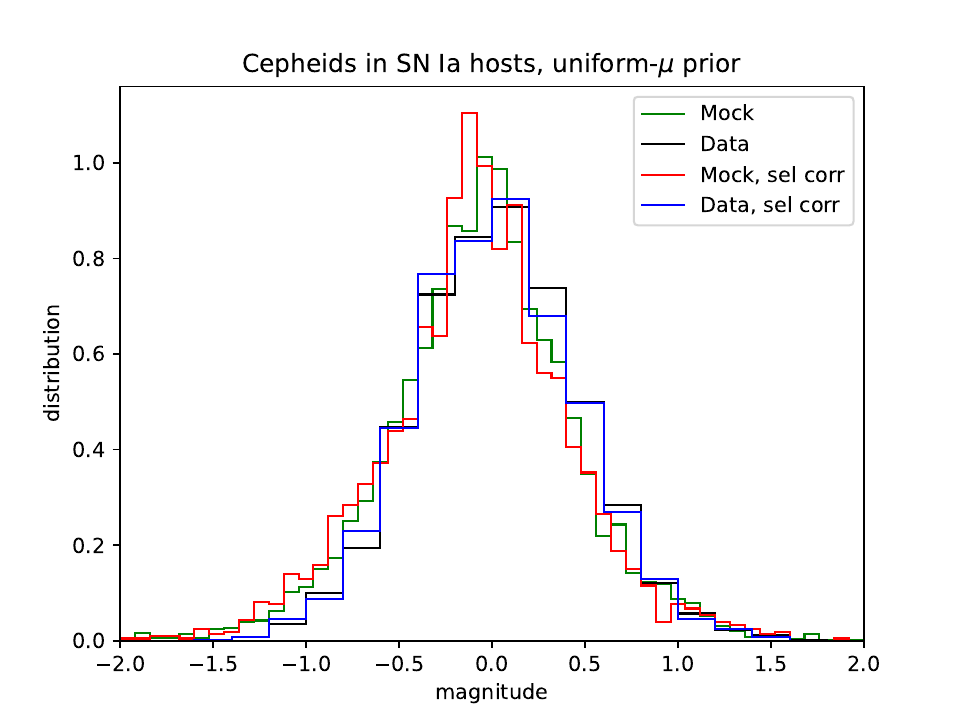}
        \includegraphics[scale=0.43]{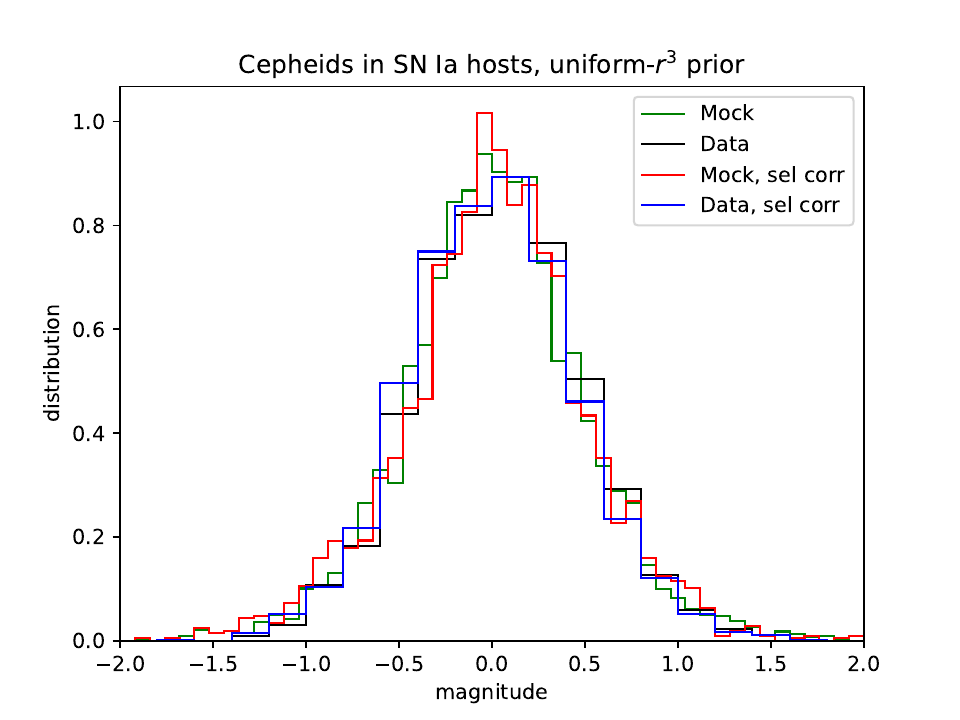}\\
        \includegraphics[scale=0.43]{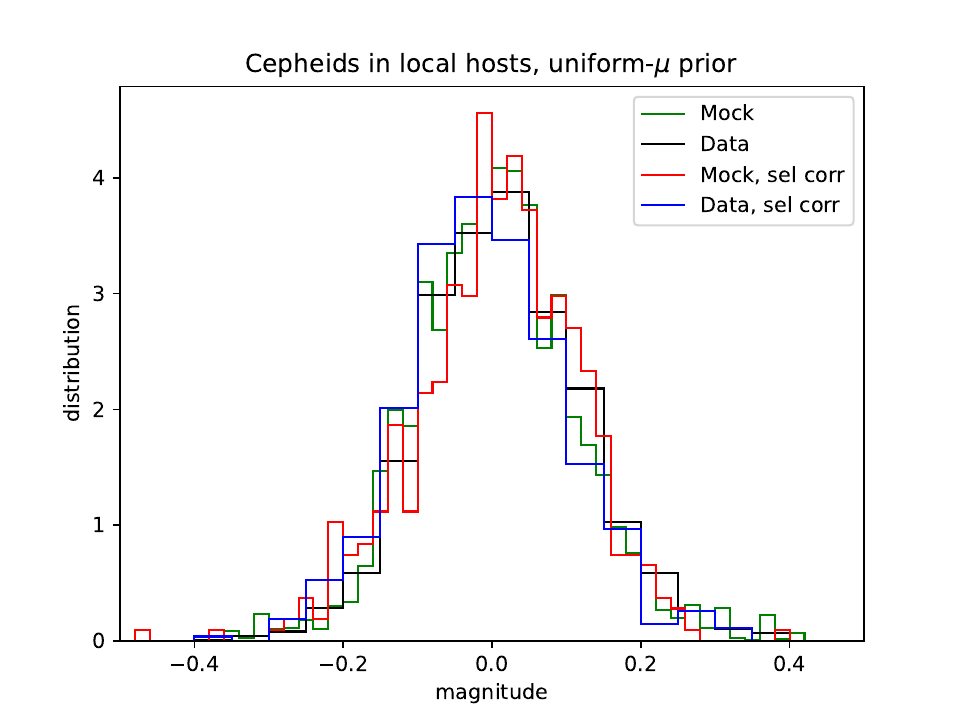}
        \includegraphics[scale=0.43]{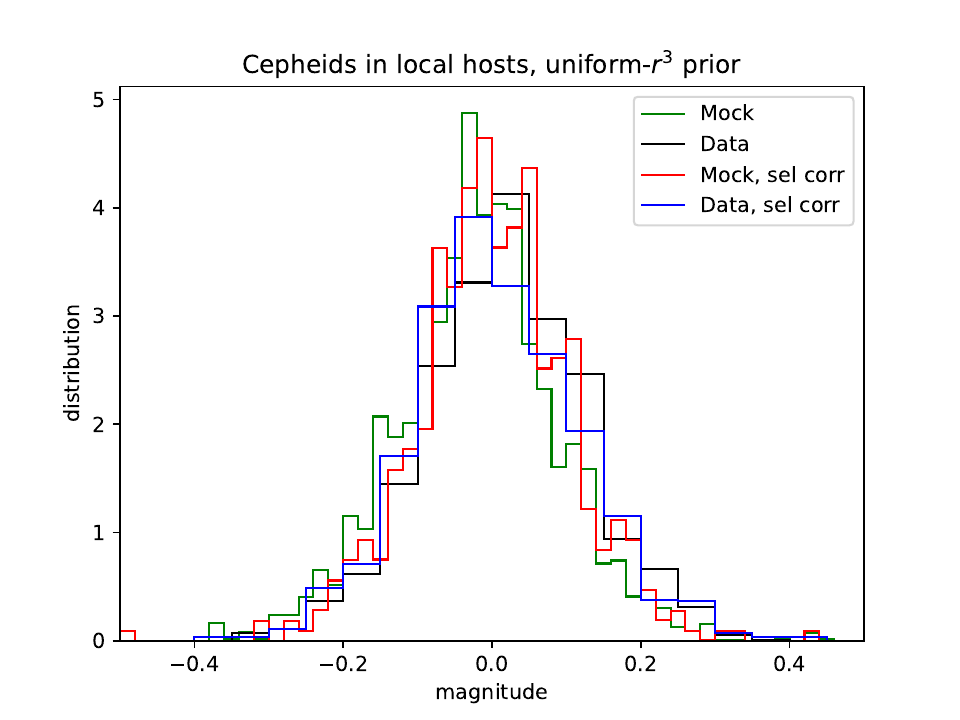}\\
        \includegraphics[scale=0.43]{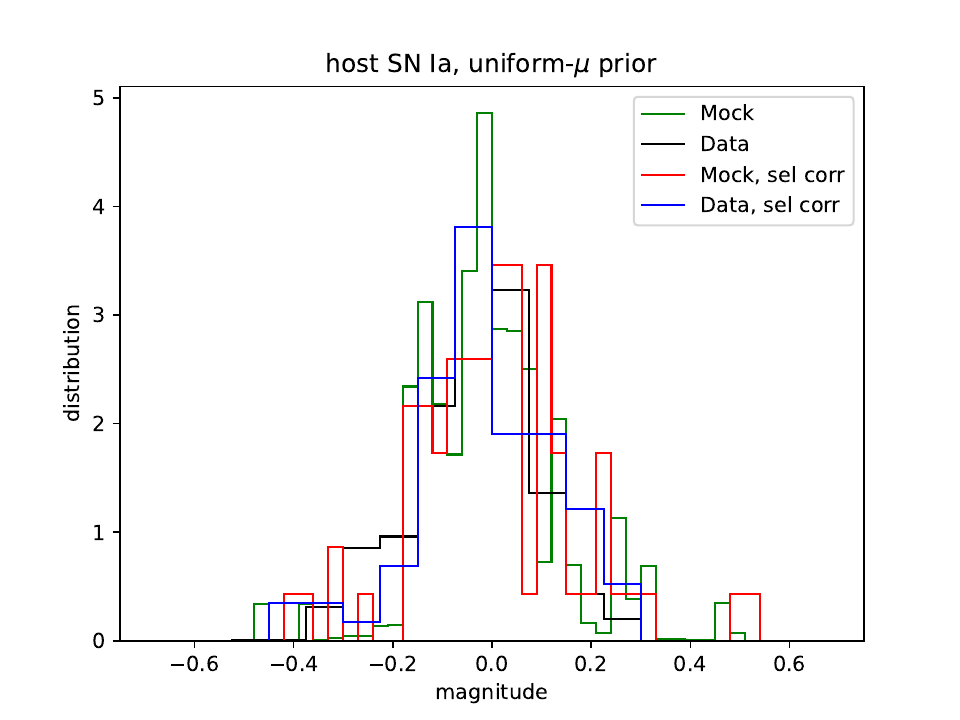}
        \includegraphics[scale=0.43]{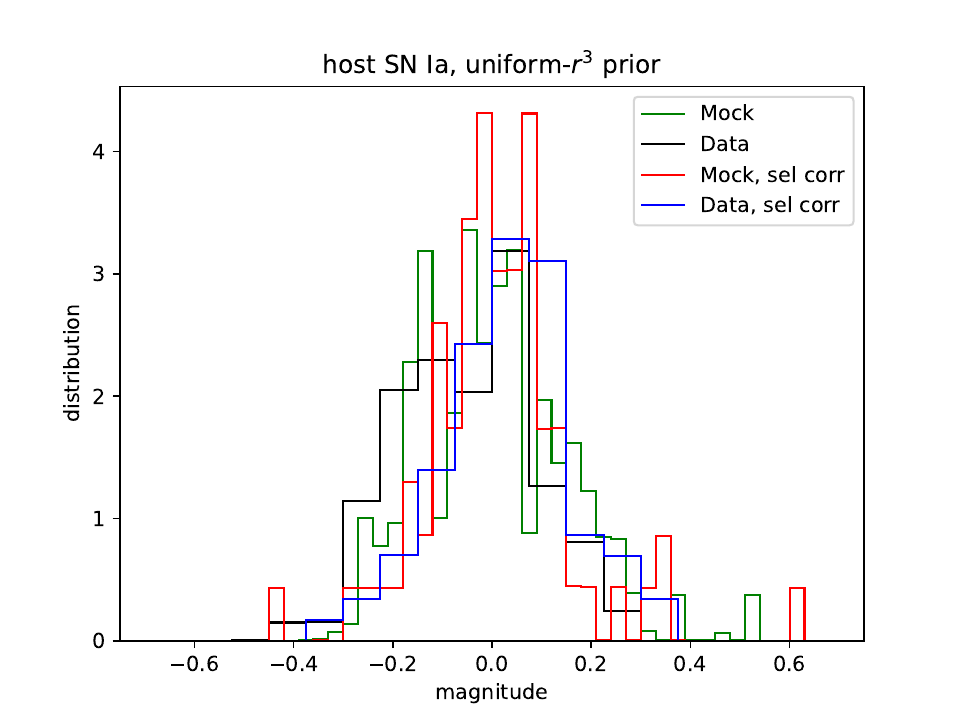}\\
        \includegraphics[scale=0.43]{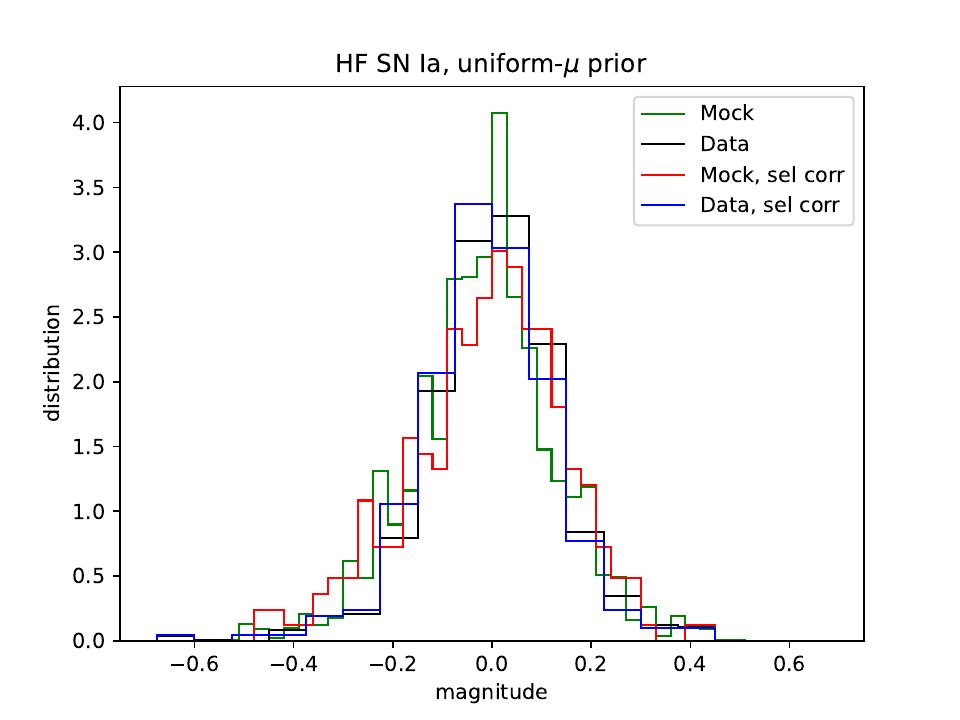}
        \includegraphics[scale=0.43]{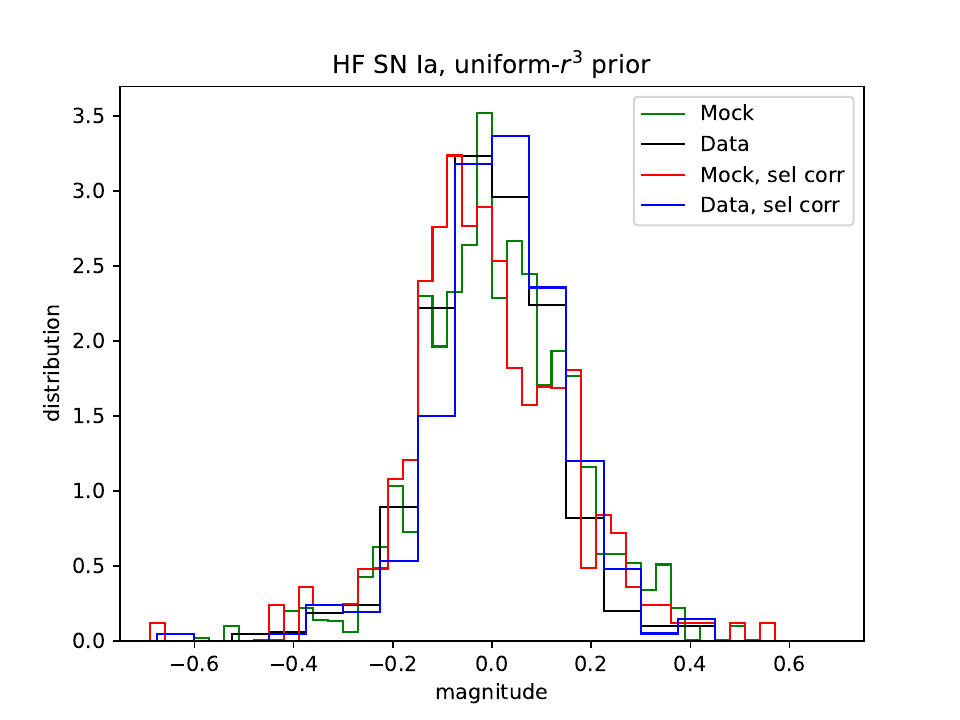}\\
    
    	\caption{Posterior predictive test (PPT) results comparing observed and simulated apparent magnitude distributions. Synthetic data sets are generated by drawing parameters from the posterior and sampling from the likelihood. We plot the distribution of offsets calculated from a set of $10^5$ simulated samples and from the measured data. We consider four subsets of magnitude measurements: Cepheids in SN Ia host galaxies and NGC 4258, Cepheids in auxiliary hosts, local SN Ia, and Hubble flow SN Ia. The left column assumes a uniform-$\mu$ prior, while the right column assumes a uniform-$r^3$ prior. Within each panel, we show results with (labelled "Corr") and without individual selection corrections applied as described in Section~\ref{sec:distance-ladder-selection}. The observed data are consistent with posterior predictive realizations under both prior choices, with no strong evidence for excess skewness or systematic offsets.}
        \label{fig:PPT}
    \end{figure}
\end{center}

\subsubsection{Magnitudes}

Our first check is on the Cepheid and SN Ia magnitudes. We split the distribution of measured magnitudes into four categories: Cepheids in SN Ia hosts and NGC 4258, Cepheids in the auxiliary hosts (M31, SMC, LMC), and local and distant SN Ia magnitudes. The distributions of magnitudes from the data and the PPT are compared in Fig.~\ref{fig:PPT}. For this comparison, we use the baseline sample from Table~\ref{table:H0-results2} including the auxilliary hosts (with results given in the penultimate row in that Table), and contrast against including no selection corrections. Visually, there is no evidence for systematic differences between the observed and predicted distributions. The mean magnitude of each sample lies comfortably within the posterior predictive distribution, whether or not we include a correction for selection effects. The most significant deviations are at the $\sim1\sigma$ level for Cepheids in SN Ia host galaxies, with no consistent pattern of increased discrepancy when selection corrections are included. We conclude that the distribution of magnitudes in the data is consistent with the assumptions underlying the Bayesian model. 

\subsubsection{Goodness of fit}

For the R22 analysis with no selection corrections, there are $3489$ data points and $46$ model parameters. Fitting to the data using this model and with uniform-$\mu$ prior, we find $\chi^2=3552.9$ between data and MAP model with $3443$ degrees-of-freedom (DoF), consistent with the possibility of mildly overestimated uncertainties. Introducing a single selection effect for extragalactic Cepheids lowers $\chi^2$ by $1.4$, while introducing $38$ parameters $s_i$ reduces $\chi^2$ by $23.3$, indicating a modest improvement in goodness-of-fit relative to the increase in model complexity. $\chi^2$ values and variation with the inclusion of selection effect corrections are similar under the uniform-$r^3$ prior. Including all selection effect corrections and the auxiliary hosts (the penultimate entry in Table~\ref{table:H0-results2}), we find a reduction in $\chi^2$ to $3494.5$ for uniform-$\mu$ priors respectively for $3398$ DoF. Removing the auxiliary hosts (M31, LMC and SMC), matching the final entry in Table~\ref{table:H0-results2}, $\chi^2=2913.6$ for $2862$ DoF. Results with the uniform-$r^3$ prior and S26 MW zero-point are similar.

We now perform a PPT to assess the Bayesian model using the $\chi^2$ statistic. We compare (i) the $\chi^2$ calculated between the data and models drawn randomly from the posterior against (ii) the $\chi^2$ calculated between the model draw and a draw from the likelihood around it. In general, the distribution from (i) will be narrower than that from (ii) because the model is trained on the same data it is tested against. For (ii), we find that the average $\chi^2$ is very close to the DoF for all models, as is to be expected from likelihood draws, with a wide range of recovered values. These encompass the $\chi^2$ values recovered from (i), which is narrower in width and is generally about $\sim50$ larger than the DoF, approximately matching the difference between $\chi^2$ and DoF recovered directly between data and MAP model. This behaviour is stable across choices of prior and inclusion of selection corrections. Given that no large discrepancies are seen for any of the model combinations presented in Tables~\ref{table:H0-results1} or~\ref{table:H0-results2} we conclude that the distribution of $\chi^2$ is consistent with the assumptions underlying the Bayesian model. 

\begin{center}
    \begin{figure}[!t]
        \centering
    	\includegraphics[scale=0.7]{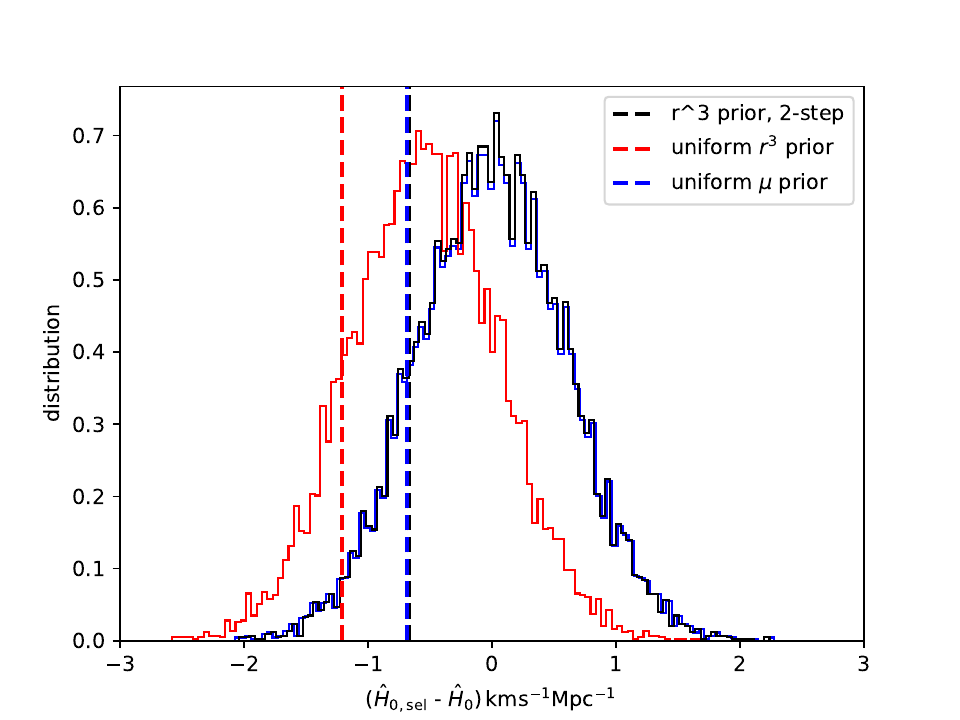}
    	\caption{The difference between the MAP values of $H_0$ obtained from fits with and without galaxy-specific selection corrections, $\Delta H_0=\hat{H}_{0,{\rm sel}}-\hat{H}_0$. The histograms show the results of $10^5$ posterior predictive simulations generated under the null hypothesis of no residual selection effects, while the dashed vertical lines indicate the values inferred from the data. MAP values of $H_0$ were calculated analytically as described in the text from real and simulated data. The blue histogram corresponds to the uniform-$\mu$ prior and is centred close to zero, indicating that the inclusion of the selection-correction parameters does not systematically shift $H_0$. In contrast, adopting a uniform-$r^3$ prior shifts the distribution away from zero, reflecting a coupling between the uniform-$r^3$ prior and the selection-correction parameters. The green histogram shows the distribution obtained using the two-step procedure described in Appendix~\ref{app:2-step}, which removes this offset at the cost of reduced Bayesian rigour.}
        \label{fig:Null-tests}
    \end{figure}
\end{center}

\subsubsection{The effect of extragalactic Cepheid selection corrections on the Hubble constant} \label{sec:null}

We now consider the effect of the selection correction included for extragalactic Cepheids on the recovered value of $H_0$. Our null test is based on the PPT principle, and we draw $10^5$ posterior samples and generate synthetic data under the likelihood with no correction, in effect creating fake data without selection effects. For each synthetic realization we compute the difference in the analytically-calculated MAP $H_0$ with and without selection correction, $\Delta H_0=\hat{H}_{0,{\rm sel}}-\hat{H}_0$, using the shared parameter model. Other than the extragalactic Cepheids, we do not include any other corrections for selection effects. This null test probes whether using the selection-augmented model of Section~\ref{sec:selection} for extragalactic Cepheids introduces a bias in $\hat{H}_0$.

Fig.~\ref{fig:Null-tests} shows the results of this test. As can be seen, for the uniform-$\mu$ prior, the posterior predictive distribution is centred near zero, indicating that introducing the selection parameters does not systematically shift $H_0$ in the absence of selection effects. Here the prior is linear and symmetric so this is expected. For the data, we find $\Delta H_0 = -0.7\,\kmsMpc$, approximately a $1\sigma$ deviation from the null distribution as discussed in Section~\ref{sec:results1}. The uniform-$r^3$ prior couples to the selection-correction parameters so, while the data and mean recovered $\hat{H}_0$ from the simulations remain consistent at $1\sigma$, the recovered difference increases to $\Delta H_0 = -1.2\,\kmsMpc$ for the data. The difference between the shifts in $H_0$ recovered from the data with the different priors matches that recovered from the mock samples in the null test. 

This behaviour arises naturally in the joint posterior when both the distance prior and selection model are included. If both are regarded as physically meaningful components of the model, then the joint posterior naturally reflects their combined influence. In Appendix~\ref{app:2-step} we consider a two-step process to correct and then fit the data, which removes the coupling but is not equivalent to inference under a single joint Bayesian model. Fig.~\ref{fig:Null-tests} shows that this can remove the coupling.

\begin{center}
    \begin{figure}[!t]
        \centering

        \includegraphics[scale=0.45]{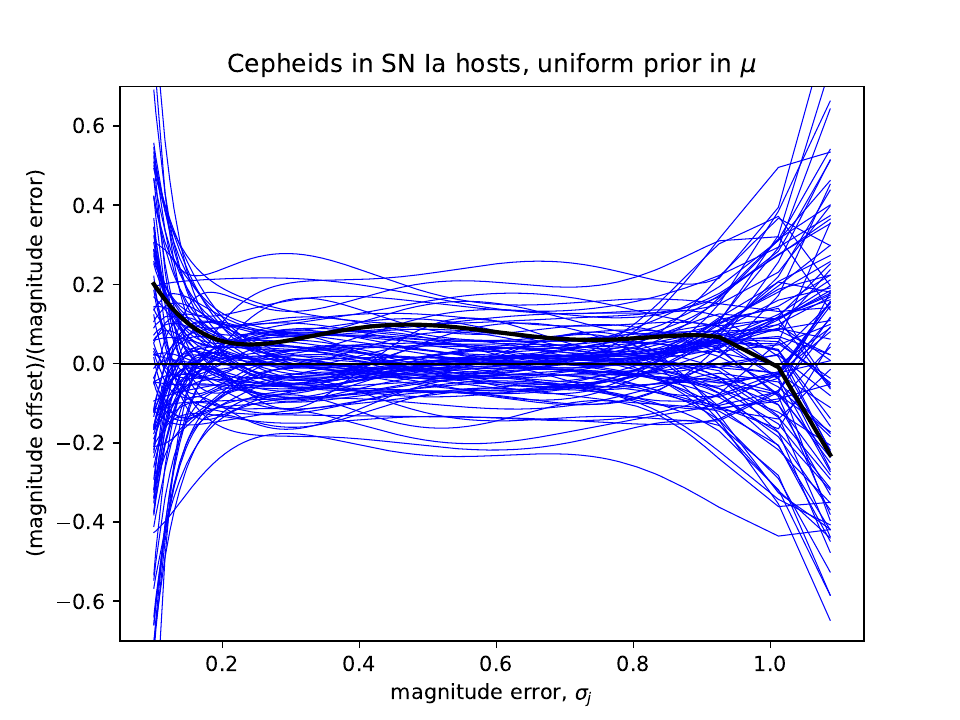}
        \includegraphics[scale=0.45]{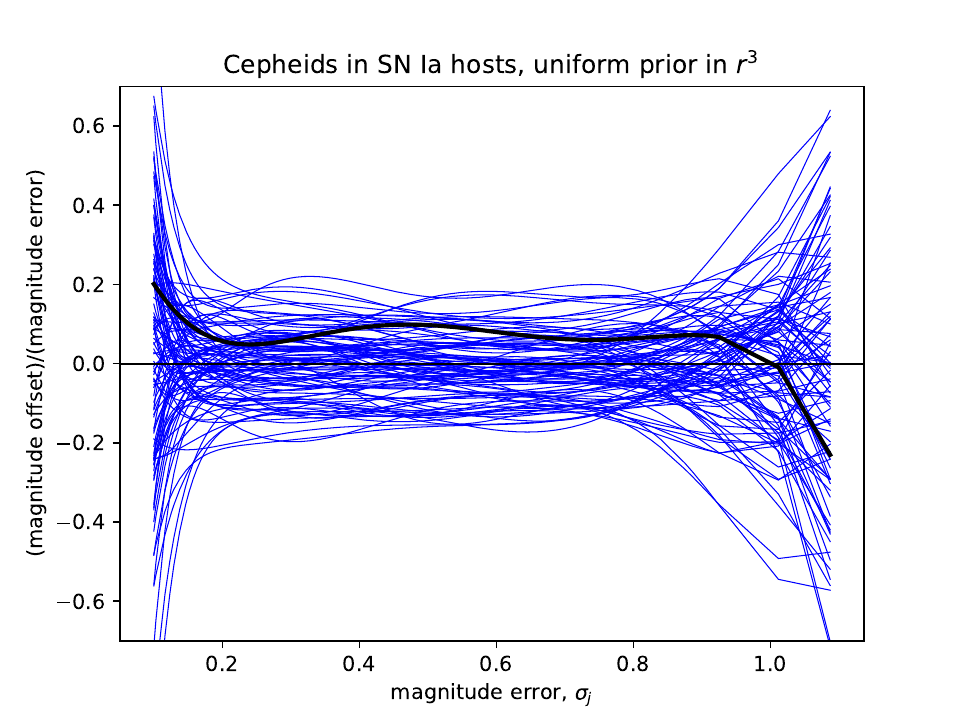}\\
        
    	\caption{Posterior predictive test (PPT) results comparing the amplitude of the extragalactic Cepheid selection correction in the MAP model. We plot the distribution of offsets for both a set of $10^4$ simulated samples (blue lines) and the actual data (thick black line). We only plot data for the Cepheids in SN Ia host galaxies and NGC 4258, as this is the most interesting sample for the correction of Section~\ref{sec:selection}. Synthetic data sets are generated by drawing parameters from the posterior from MCMC runs with no correction terms included, and then sampling from the likelihood. The left panel assumes a uniform-$\mu$ prior, while the right panel assumes a uniform-$r^3$ prior. The observed data are consistent with posterior predictive realizations under both prior choices, as expected given the weak evidence for a correction term that scales with the magnitude error, $\sigma_j$. The evidence is discussed further in Section~\ref{sec:significance}.}
        \label{fig:PPT-off}
    \end{figure}
\end{center}

\subsubsection{Null test on the amplitude of the Cepheid selection corrections} \label{sec:offset}

Our final PPT test is on the amplitude of the Cepheid selection correction terms in the model. We draw $10^2$ posterior samples from MCMC runs with no correction terms included, and generate synthetic data under the likelihood. These realizations are not affected by selection effects. For each synthetic realization and for the data we analytically calculate the MAP solution including selection terms $s_i$ for the $38$ SN Ia hosts and NGC4258. The selection correction term is then calculated as the difference between the MAP model and the MAP model when the $s_i$ are set to zero. The selection corrections required are plotted in Fig.~\ref{fig:PPT-off} as a function of the magnitude error $\sigma_j$. 

We see that the data fall within the range of simulations, as expected given the weak evidence of selection effects as described in Section~\ref{sec:significance}. Nevertheless, the data lie within the upper range of results from the synthetic spectra. The significance of requiring non-zero $s_i$ is considered in Section~\ref{sec:significance}. From Fig.~\ref{fig:PPT-off}, we also see that, even though the $s_i$ are allowed to vary between galaxies, the offset is proportional to $\sigma_j$.

\section{Discussion}
\label{sec:concl}

A simple model for selection effects in data correlated with that used to make measurements was introduced in Section~\ref{sec:selection}. While clearly basic, the model illustrates a generic statistical phenomenon rather than attempting to reproduce the full selection function of any sample. It is based on the fundamental concept, common in astronomy, of truncation-limited selection (e.g. on signal-to-noise) and Normal errors, leading to a sample of selected objects. The properties of the detected objects are then measured using correlated observations. This can lead to a sample whose likelihood shape does not reveal selection effects, but that has an offset in the mean from a complete sample. Provided that the selection criteria and correlation coefficient are constant, the size of the offset is dependent on the measured error. It is thus possible to measure the selection effect from data, given a range of measurements with different recovered errors, provided the correlation and cut-off are the same for all objects.  

We investigate whether residual selection effects of this type remain in the extragalactic Cepheid sample included in the distance ladder fit of R22. This model is appropriate because the R22 analysis selects Cepheids in distant galaxies using different observations from those used to measure their magnitudes, and we expect correlations between the two from correlated background fluctuations across wavebands. However, given the difficulty in modelling the selection, we cannot be certain that the simple model is sufficiently well matched to the actual selection effects. Nevertheless it is interesting to investigate the impact of such an effect. Our method of assessing the significance of detection by calculating the weighted offset of the selection parameters from zero gives the same result whether fitting a shared parameter for the size of the selection effect or individual parameters for each Cepheid host. We find a slight preference for non-zero selection parameters ($1.3$ to $2.3\sigma$ for different priors). 

Although these significances do not constitute compelling evidence for residual selection effects, nor do they demonstrate that residual selection effects can safely be neglected when assessing the robustness of $H_0$ to the selection of Cepheids in nearby galaxies. We therefore regard it as informative to explore the impact on $H_0$ when these corrections are marginalized over. Including a shared parameter to control the level of selection effect for all galaxies, we find that $H_0$ is reduced by $\Delta H_0=-0.8\,\kmsMpc$ or $\Delta H_0=-1.4\,\kmsMpc$, for the uniform-$\mu$ and uniform-$r^3$ distance priors respectively. We have not performed Bayesian model comparison \cite[e.g.][]{Paradiso2024} to determine the requirement for a model with Cepheid selection corrections because Bayes factors are highly prior sensitive for weakly constrained parameters. The parameters are weakly constrained, which we know already from the fits. Therefore Bayes factors are not especially informative. 

The results are robust to priors placed directly on key nuisance parameters in Milky Way Cepheid fits ($M_{H}^W$) and on $H_0$ itself. In contrast, the inference shows a sensitivity to priors on distance parameters. For Milky Way Cepheids, replacing a uniform prior in distance modulus with a spatially motivated prior based on the Galactic stellar distribution \citep{BailerJonesetal2021} rather than a uniform prior in $\mu$ reduces $H_0$ by $\sim0.8\,\kmsMpc$, which is approximately cancelled by the correction for selection effects in this sample \citet{Stiskalek2026}. Extending this class of priors to extragalactic distances by adopting a uniform-$r^3$ prior further reduces $H_0$ relative to a uniform-in-distance-modulus prior, as used in R22. Together, these changes produce a net shift of $\sim-1.3\,\kmsMpc$, consistent with the results of \citet{Desmond2025DistancePrior,hogas2026physically}, when corrected for the selection effects as in \citet{Stiskalek2026}.

We have also considered adjusting the likelihood to include selection effects for the SN Ia samples included in the distance ladder. For the extragalactic Cepheid host galaxies, we adopt the simple selection model of \citet{Stiskalek2025}, while for the Hubble flow SN Ia, we again use the model of Section~\ref{sec:selection} to test for additional selection corrections of this form. Consistently using the uniform-$r^3$ prior and the MW zero-point of \citet{Stiskalek2026}, we recover $H_0=70.23\pm1.72\,\kmsMpc$. The selection of the Cepheids included in R22 in the auxilliary hosts, M31, the SMC and LMC is more complicated and we do not attempt to include it here. As a robustness test, removing these data gives $H_0=69.76\pm1.96\,\kmsMpc$. We caution, however, that simply removing data and presenting an increased error bar does not resolve the Hubble tension.

It is intriguing that the attempt to include selection effects for all of the samples used in the distance ladder presented here has altered the recovered marginal value of $H_0$ by a shift that is large compared to the quoted statistical error from distance ladder measurements (\citealt{Casertano2026} found an error bar on $H_0$ of $\pm\,0.81\,\kmsMpc$). The offset from R22 matches the sum of shifts from applying selection effect corrections and distance priors and a small term arising from the coupling of these effects. This coupling is strongest for the Cepheid selection, and we saw that allowing individual corrections for each host increases the shift beyond the sum of the individual effects because of the increased freedom allowed by that fit (see Appendix~\ref{app:2-step}). It is interesting that a weakly constrained selection model can interact non-trivially with the choice of distance prior, producing an $H_0$ shift that is not simply the sum of the  individual effects. We see no evidence for such a coupling for the Hubble flow SN Ia, with the shifts seen including this effect in Table~\ref{table:H0-results2} being consistent for both priors.

Useful future work includes forward-modelling joint extragalactic Cepheid selection and photometric measurement processes, providing a more direct connection between selection assumptions and the underlying observational process. Because the predicted bias scales with the measurement uncertainty, improved photometric precision will alter the magnitude of this class of selection effect, although the correlation coefficient will also change, so the value of ${\bf s}$ will be different. Thus, the program of work currently underway by the SH$0$ES team to obtain JWST photometry of extragalactic Cepheids \citep{Riess2024,Riess2025} will provide an important test of the inferred selection corrections. Current JWST measurements give errors of order $0.03$\,mag, compared to typical HST magnitude error of $0.38$\,mag per Cepheid: we would expect this class of selection effect to become less important if the relevant correlations and selection mechanisms remain comparable. Current data are, however, insufficient, given the small number of galaxies observed, as evidenced by the scatter seen in Fig.~\ref{fig:Delta_m}. Further posterior predictive tests such as LOO-PIT \citep{LOO-PIT} would test the robustness of individual measurements, galaxies and Cepheids within the analysis. Further investigation into the selections is warranted for all of the samples given the importance demonstrated here and the assumptions made in this work. Finally, independent distance ladder analyses will be important for cross-validation of these modelling sensitivities \citep{Casertano2026}. 

More generally, the formalism developed here provides a framework for incorporating correlated-observable selection effects into Bayesian inference. Although illustrated using extragalactic Cepheid selection, the same statistical phenomenon can arise whenever sample selection is based on one observable and inference is performed on another correlated observable. 

\section*{Acknowledgements}

WJP acknowledges useful conversations with Marcus H\"og{\aa}s, Dustin Lang, Ana Ennis, Marco Bonici, and George Efstathiou, and thanks Adam Riess for extensive discussions on an early draft. AI was used to support the coding used in this project and writing the paper, at the level of providing simple routines and correcting unclear language and descriptions: WJP takes full responsibility for the result. Support was provided by the Natural Sciences and Engineering Research Council of Canada (NSERC), [funding reference number RGPIN-2025-03931] and from the Canadian Space Agency. This research was enabled in part by support provided by Compute Ontario (computeontario.ca) and the Digital Research Alliance of Canada (alliancecan.ca).

\bibliographystyle{apsrev4-1}

% You should give the same name for your .bbl as your main .tex
% since it is a requirement for posting on ArXiv.
\bibliography{H0-distance-ladder}

\begin{appendix}
\section{two-step approach to including Cepheid selection corrections}
\label{app:2-step}
\begingroup 
    \setlength{\tabcolsep}{10pt} % Default value: 6pt
    \renewcommand{\arraystretch}{1.5} % Default value: 1
    \setlength\extrarowheight{2pt}
    \begin{table}
        \centering
        \begin{tabular}{ c c c }  
            selection & \multicolumn{2}{c}{$H_0\,\kmsMpc$} \\
            model & uniform-$\mu$ & uniform-$r^3$ \\
            \hline \hline
            R22, no CV selection & $H_0=73.04\pm1.01$ & $H_0=72.04\pm0.99$ \\ % run 00100000, 10100000
            R22, CV individual & $H_0=72.24\pm1.33$ & $H_0=70.05\pm1.29$ \\ % run 00110000, 10110000
            R22, CV two-step & $H_0=72.21\pm1.00$ & $H_0=71.22\pm0.99$ \\ % run 00310000, 10310000
            \hline \hline  
        \end{tabular}  
        \caption{Recovered values of the Hubble constant under different corrections for extragalactic selection effects. We contrast results with no correction, with a correction based on a full Bayesian fit, and the two-step method designed to test the coupling between prior and correction.}
        \label{table:H0-results3}
    \end{table}
\endgroup

To investigate the coupling of prior and selection correction, we consider a two-step approach, that corrects and then fits the data. To test this, we adopt the model with individual $s_i$ for each Cepheid host, as this was most affected by the distance prior, suggesting the largest coupling. The first step is to determine the posterior mean $\hat{s}_i$ using Eq.~\ref{eq:best-fit} with uniform priors on the distance moduli and $s_i$. These are then used to correct the data. The corrected data are then fit using the original 46-parameter model. In principle, obtaining results identical to the full parameter fit requires propagating the uncertainty from the inferred selection corrections through to the covariance matrix. This can be achieved using $PCP^T$, where the propagator matrix $P=I-W(L'C^{-1}L'^T)^{-1}L'C^{-1}$ maps the data vector onto its selection-corrected counterpart ${\bf y}-\hat{s}_i\sigma_{j}$. Instead, to decouple the selection corrections from the distance priors, we can treat the corrections as fixed and retain the original covariance matrix $C$. This procedure does not provide correct parameter uncertainties as it does not allow for coupled errors in $\hat{s}_i$. Results are presented in Table~\ref{table:H0-results3}. As expected, the two-step correction results with a uniform-$r^3$ prior are consistent with a linear combination of the shift due to prior coming from the uniform$-\mu$ results and selection correction when either is introduced separately.

\end{appendix}

\end{document}